\documentclass[10pt]{iopart}
\usepackage{iopams}
\expandafter\let\csname equation*\endcsname\relax
\expandafter\let\csname endequation*\endcsname\relax
\usepackage{amsmath,amsfonts,amssymb,amsthm}
\usepackage{amsmath,amsfonts,amssymb,amsthm}
\usepackage{txfonts}
\usepackage{graphicx}
\usepackage{enumerate}
\usepackage{cite}
\begin{document}

\title[]{Newton's equation of motion with quadratic drag force and Toda's potential as a solvable one}

\author{Daisuke A. Takahashi}

\address{
Research and Education Center for Natural Sciences, Keio University, Hiyoshi 4-1-1, Yokohama, Kanagawa 223-8521, Japan
}
\ead{daisuke.takahashi@keio.jp}
\begin{abstract}
The family of exactly solvable potentials for Newton's equation of motion in the one-dimensional system with quadratic drag force has been determined completely. The determination is based on the implicit inverse-function solution valid for any potential shape, and hence exhaustive. This solvable family includes the exponential potential appearing in the Toda lattice as a special limit. The global solution is constructed by matching the solutions applicable for positive and negative velocity, yielding the piecewise analytic function with a cusp in the third-order derivative, i.e., the jerk. These procedures and features can be regarded as a generalization of Gorder's construction [Phys. Scr. 2015, {\bf 90}, 085208] to the energy-dissipating damped oscillators. We also derive the asymptotic formulae by solving the matching equation, and prove that the damping of the oscillation amplitude is proportional to $ t^{-1} $.
\end{abstract}

%
\vspace{2pc}
\noindent{\it Keywords}: quadratic drag force, Toda's exponential potential, elliptic functions, piecewise analytic functions 

%
\submitto{\PS}
%
%
\ioptwocol

\section{Introduction}
\indent The theory of nonlinear oscillators, while the subject itself is classical and possesses a very long history \cite{LambDynamics,WhittakerTreatiseOn}, has been still receiving renewed interest even in recent years, by finding fruitful relationships and applications in the modern scientific topics. For example, the dynamical systems with many degrees of freedom often falls into a single oscillator equation near the bifurcation point, called the normal form \cite{GuckenheimerHolmes}. The normal form predicts universal scaling laws in various physical quantities which are far from intuitive from the original equation; the examples of scaling behaviors in the nonlinear Schr\"odinger equation with a spatial inhomogeneity are found in Refs. \cite{PhysRevE.55.2835,HUEPE2000126,PHAM2002127,TAKAHASHI20121589,PhysRevLett.105.035302} and applied to the superfluidity phenomena in Bose-Einstein condensates. As another important aspect, the nonlinear oscillators play a role as a testbed to verify modern approximation techniques such as the renormalization group perturbation theory \cite{PhysRevE.54.376,doi:10.1143/PTP.94.503}, homotopy analysis/perturbation methods \cite{doi:10.1142/S0217979206033796,LiaoHAM}, and so on. \\
\indent In the light of the above-mentioned applications, the solvable models with closed-form solutions could be a powerful tool, since the accuracy of the numerical solution can be tested by directly comparing with the exact one. 
In particular, if the system is described by the equations and/or solutions \textit{without} fundamental mathematical analyticity, the existence of the exact solution becomes further important, because it is not evident whether a number of general mathematical theorems relying on good differentiability are applicable or not. Gorder's work \cite{1402-4896-90-8-085208}, in which the non-differentiable particle motion under the logarithmic potential has been derived, is an illustrative example. As explained below, the problem addressed in this paper also belongs to the same category. \\
\indent Here, we clarify the problem which we solve in this paper. Let us consider the following one-dimensional and one-particle equation of motion with the resistive force proportional to the velocity squared (the quadratic drag force):
	\begin{align}
		M \ddot{x}+\gamma (\operatorname{sgn}\dot{x})\dot{x}^2+U'(x)=0, \label{eq:eom}
	\end{align}
	where $ M>0 $ is the mass,  $ \ddot{x} $ is the acceleration, $ \dot{x} $ is the velocity, $ \gamma>0 $ is the drag coefficient, and  $ U(x) $ is the potential. Such a drag force becomes important when an object in the air has a high speed. Then, the aim of the present paper is phrased as follows: \textit{Identify the family of solvable potentials $ U(x) $ for Eq.~(\ref{eq:eom}) exhaustively, and constructing the explicit solution, elucidate its analytic and asymptotic natures.} \\
	\indent The sign function $ \operatorname{sgn}\dot{x} $ in Eq.~(\ref{eq:eom}) is in order for the drag force to have an opposite direction to the motion of the particle, and due to this term, the differential equation (\ref{eq:eom}) is not analytic at $ \dot{x}=0 $. Hence, if we are interested in an exact solution, we must solve it for the cases $ \dot{x}\gtrless 0 $ separately, and join the solutions smoothly. 
	The solution for the uniform gravity $ U(x)=Mgx $ can be found in numerous textbooks \cite{LambDynamics,WhittakerTreatiseOn}, but we will provide a richer family which are solvable within the elementary and elliptic functions. This family includes Toda's exponential potential appearing in the Toda lattice \cite{Todalattice} as a special limit. We especially focus on this special case and construct the solutions with their asymptotics. \\
\indent The recent relevant studies in similar or modified settings include the pseudo-oscillator equation with the logarithmic potential modeling the path of electrons in a plasma tube \cite{1402-4896-89-10-105205,1402-4896-90-8-085208}, the equation without $ \operatorname{sgn}\dot{x} $ describing the granular materials \cite{PhysRevE.51.2538,0143-0807-16-2-004} and those with higher-order potentials \cite{1402-4896-85-4-045006}, the projectiles in two dimension \cite{doi:10.1119/1.10812,1751-8121-40-29-015}, the relativistic and Duffing oscillators governed by homotopy methods \cite{1402-4896-77-2-025004}, and the Duffing oscillator with linear damping force \cite{0143-0807-36-6-065020}. In particular, Ref.~\cite{1402-4896-90-8-085208} shares a common feature with our present work, since the solution is constructed by matching procedure and the solution becomes singular at the matching point. In our work, we demonstrate that the similar matching method can be applicable even for the energy non-conserving damped system, where the value of the integration constant gradually changes at every matching point, and the cusp appears in the third-order derivative, that is, the jerk. \\
 	\indent The organization of the paper is as follows. In Sec.~\ref{sec:implicitsol}, we derive a general implicit solution applicable for any potential. In Sec.~\ref{sec:finidtoda}, we find a family of solvable potentials including Toda's one.  In Secs.~\ref{sec:diffsol} and \ref{sec:connectingsols}, we construct the solution for Toda's potential. We solve the differential equation for the cases $ \dot{x}\gtrless0 $, and matching them smoothly, we construct the global solution. In Sec.~\ref{sec:asym}, we provide an asymptotic behavior of the solution. Section~\ref{sec:summ} is devoted to the discussion, summary, and future outlook.

\section{Implicit solution for general $ U(x) $}\label{sec:implicitsol}
We first derive an implicit solution to Eq.~(\ref{eq:eom}) for general $ U(x) $. 
Since Eq.~(\ref{eq:eom}) does not include $ t $ explicitly, if $ x(t) $ is a solution,  $ x(t-t_0) $ is also a solution. Therefore, if we rewrite the equation with respect to not $ x(t) $ but its inverse function $ t(x) $,  it contains only  $ t'(x) $ and $ t''(x) $. Defining $ s(x):=t'(x) $, and using the formulas for inverse functions $ \dot{x}=1/s $ and $ \ddot{x}=-s'/s^3 $, the resultant equation is
	\begin{align}
		-Ms'+\gamma |s|+s^3U'(x)=0.
	\end{align}
	Thus, the second-order equation is reduced to the first-order one. This equation can be linearized by substituting $ s=\pm f^{-1/2} $ for $ s \gtrless 0 $, and we obtain the solution in an implicit form
	\begin{align}
		t= \sqrt{\frac{M}{2}} \int \frac{\pm\mathrm{e}^{\pm\kappa x/2}dx}{\sqrt{C-\int \mathrm{e}^{\pm\kappa x}U'(x)dx}},\quad \kappa:=\frac{2\gamma}{M}. \quad (\dot{x}\gtrless0) \label{eq:formalsol}
	\end{align}
	If $ \kappa=0 $, the physical interpretation of the constant $ C $ becomes the energy $ E $, and the solution reduces to the well-known form $ t=\sqrt{\frac{M}{2}}\int[E-U(x)]^{-1/2}dx $ (e.g., chapter III of Ref.~\cite{LLMechanics}). \\
	\indent The solution (\ref{eq:formalsol}), though it is applicable for any $ U(x) $, is not actually so convenient, unless the solution can be obtained in an explicit form ``$x(t)=\dots$''. The reason is as follows.  In the energy-conserving system (i.e., $ \kappa=0 $), even if the integration cannot be performed explicitly for general $ U(x) $, it is still useful to identify the region where the particle motion is possible for a given $ E $. Furthermore, if the solution is given by a periodic oscillation, the global solution for fixed $ E $ can be constructed by repeating the copy-and-paste of the one-period solution. On the other hand, in the present system, we need to construct a global solution by connecting the solutions for $ \dot{x}\gtrless0 $, where the value of the constant $ C $ changes at every junction point $ \dot{x}=0 $ because of the damping by drag force. Hence, the solution cannot be obtained by a naive repetition of the one-period solution.  Therefore, finding a potential such that the explicit closed-form solution is available for arbitrary $ C $ is essential.
	
\section{Finding a solvable potential}\label{sec:finidtoda}
	Let us now find a solvable potential such that the integration in Eq.~(\ref{eq:formalsol}) can be performed explicitly. 
	Introducing the new variables $ z=\mathrm{e}^{\pm\kappa x} $ for the cases $ \dot{x}\gtrless0 $, Eq.~(\ref{eq:formalsol}) is rewritten as
	\begin{align}
		t=\begin{cases} \displaystyle \sqrt{\frac{M}{2\kappa}}\int^{\exp(\kappa x)} \frac{dz}{\sqrt{z\left[ C\kappa -\int u(z)dz \right]}} & (\dot{x}>0), \\ \displaystyle \sqrt{\frac{M}{2\kappa}}\int^{\exp(-\kappa x)} \frac{dz}{\sqrt{z\left[ C\kappa+\int u(z^{-1})dz \right]}} & (\dot{x}<0), \end{cases} \label{eq:formalsol2}
	\end{align}
	where $ u(z):= U'\left(\frac{\ln z}{\kappa}\right) $.  We now seek a potential such that Eq.~(\ref{eq:formalsol2}) can be calculated within the elementary and elliptic functions. It imposes the condition that \textit{both $ z\left[ C\kappa -\int u(z)dz \right] $ and $ z\left[ C\kappa+\int u(z^{-1})dz \right] $ are polynomials of order $ \le $ 4}, and it fixes the form of the potential as $ u(z)=a+bz^2+cz^{-2} $. The corresponding $ U(x) $ is given by
	\begin{align}
		U(x)=\frac{A}{2\kappa}\left( 2\kappa x\tanh2\kappa d+\frac{\cosh 2\kappa(x-d)}{\cosh 2\kappa d}-1 \right)  \label{eq:familypot}
	\end{align}
	with $ A \in \mathbb{R} $ and $ d\in \mathbb{R}  $ or $ \mathbb{R}+\frac{\mathrm{i}\pi}{4\kappa} $. Here, we chose $ U(0)=U'(0)=0 $ without loss of generality. In this normalization,  $ u(z) $ is written as
	\begin{align}
		u(z)=A\left( \frac{z_0^{-2}z^2-z_0^2z^{-2}}{z_0^2+z_0^{-2}}+\frac{z_0^2-z_0^{-2}}{z_0^2+z_0^{-2}} \right)
	\end{align}
	with $ d=\frac{\ln z_0}{\kappa} $. \\
	\indent If $ d \in \mathbb{R} $ and $ A>0 $, the potential (\ref{eq:familypot}) is non-negative everywhere and the particle motion is always bounded. If $ d \in \mathbb{R}+\frac{\mathrm{i}\pi}{4\kappa} $, Eq.~(\ref{eq:familypot}) has local extrema at $ x=0 $ and $ 2d $, and the motion may not be bounded depending on the initial condition. If we set $ d \to +\infty $, it reduces to
	\begin{align}
		U(x)=A\left( x+\frac{\mathrm{e}^{-2\kappa x}-1}{2\kappa} \right), \label{eq:toda}
	\end{align}
	which is just the famous potential appearing in the Toda lattice \cite{Todalattice}. The emergence of Toda's exponential potential in the present context is a little unexpected, since we have made no soliton-theoretical consideration here. In this case the corresponding $ u(z) $ is given by $ u(z)=A(1-z^{-2}) $. Henceforth, we concentrate on this special potential and investigate its solutions in detail.

\section{Solutions of the differential equations for $ \dot{x}\gtrless0 $}\label{sec:diffsol}
	In this and the next section, we construct the solution for the potential (\ref{eq:toda}). 
	Among the solvable family in Eq.~(\ref{eq:familypot}), this potential is the easiest one in the sense that the solution for $ \dot{x}>0 $ becomes an elementary function. \\ 
	\indent This section provides the list of the solutions to the differential equations for $ \dot{x}\gtrless0 $, which will become the ``pieces'' of the global solution given in Sec.~\ref{sec:connectingsols}. Henceforth we follow the convention by Abramowitz and Stegun for the notation of the elliptic integrals and functions. \\
	\indent For $ u(z)=A(1-z^{-2}) $, Eq.~(\ref{eq:formalsol2}) reduces to
	\begin{align}
		t = \begin{cases}\displaystyle \sqrt{\frac{M}{2\kappa A}}\int^{\exp(\kappa x)}\frac{dz}{\sqrt{2c_+ z-1-z^2}} & (\dot{x}>0), \\ \displaystyle \sqrt{\frac{3M}{2 \kappa A}}\int^{\exp(-\kappa x)}\frac{dz}{\sqrt{z(c_-+3z-z^3)}} & (\dot{x}<0), \end{cases}
	\end{align}
	where $ c_+=\frac{\kappa C}{2A} $ and $ c_-=\frac{3 \kappa C }{A} $. There are one real-valued solution for $ \dot{x}>0 $ and three for $ \dot{x}<0 $ depending on the values of constants $ c_\pm $. Below we summarize it.  \\
	\indent I. The case $ \dot{x}>0 $. The real-valued solution exists when $ c_+ \ge 1 $. Writing $ c_+=\cosh\kappa x_0 $ with $  x_0\ge0 $, the solution is given by
	\begin{align}
		&x_{\text{I}}(t,x_0):=\frac{1}{\kappa}\ln\left[ \cosh\kappa x_0-(\sinh\kappa x_0)\cos\omega_{\text{I}} t \right], \label{eq:xI} \\
		&\omega_{\text{I}}=\sqrt{\frac{2\kappa A}{M}}. 
	\end{align} 
	\indent II. The case $ \dot{x}<0 $ and $ -2 \le c_-\le 0 $. Parametrizing $ c_-=-2\cos3\varphi,\ \varphi \in [0,\frac{\pi}{6}] $, the polynomial is factorized as $ z(c_-+3z-z^3)=-z\prod_{n=0,1,2}[z+2\cos( \varphi-\frac{2n\pi}{3} )] $. The resultant solution is then given by
	\begin{align}
		&x_{\mathrm{II}}(t;\varphi):=\frac{1}{\kappa}\ln\left[ \frac{\operatorname{sn}^2\left(\frac{\omega_{\text{II}}t}{2}\big|m_{\text{II}}\right)}{2\sin(\frac{\pi}{6}+\varphi)}+\frac{\operatorname{cn}^2\left(\frac{\omega_{\text{II}}t}{2}\big|m_{\text{II}}\right)}{2\sin(\frac{\pi}{6}-\varphi)} \right], \label{eq:xII1} \\
		&m_{\text{II}}=m_{\text{II}}(\varphi):=\frac{\sin2\varphi}{\cos(2\varphi-\frac{\pi}{6})},\\
		&\omega_{\text{II}}=\omega_{\text{II}}(\varphi):=\omega_{\text{I}}\sqrt{\frac{2\cos(2\varphi-\frac{\pi}{6})}{\sqrt{3}}}. \label{eq:xII2}
	\end{align}  
	\indent III. The case $ \dot{x}<0 $ and $ 0 \le c_-\le 2 $. Parametrization is the same as II, but $ \varphi \in [\frac{\pi}{6},\frac{\pi}{3}] $. The solution is then given by
	\begin{align}
		&x_{\mathrm{III}}(t;\varphi):=\frac{1}{\kappa}\ln\left[ \frac{\operatorname{ns}^2\left(\frac{\omega_{\text{III}}t}{2}\big|m_{\text{III}}\right)}{2\sin(\frac{\pi}{6}+\varphi)}-\frac{\operatorname{cs}^2\left(\frac{\omega_{\text{III}}t}{2}\big|m_{\text{III}}\right)}{2\sin(\frac{\pi}{6}-\varphi)} \right],\\
		&m_{\text{III}}=m_{\text{III}}(\varphi):=\frac{\cos(2\varphi-\frac{\pi}{6})}{\sin2\varphi},\\
		&\omega_{\text{III}}=\omega_{\text{III}}(\varphi):=\omega_{\text{I}}\sqrt{\frac{2\sin2\varphi}{\sqrt{3}}}.
	\end{align}
	\indent IV. The case $ \dot{x}<0 $ and $ 2 \le c_- $. Parametrizing $ c_-=2\cosh3\varphi,\ \varphi\ge0 $, the solution is  
	\begin{align}
		 &x_{\mathrm{IV}}(t;\varphi):=\nonumber \\
		 &\frac{1}{\kappa}\ln \left[ \frac{1}{2\cosh\varphi} \left( 1+\!\!\sqrt{\frac{3(2\cosh2\varphi+1)}{2\cosh2\varphi-1}}\frac{1-\operatorname{cn}(\omega_{\text{IV}}t|m_{\text{IV}})}{1+\operatorname{cn}(\omega_{\text{IV}}t|m_{\text{IV}})} \right) \right],\\
		 &m_{\text{IV}}=m_{\text{IV}}(\varphi):=\frac{1}{2}-\frac{\sqrt{3}}{2}\frac{\cosh2\varphi}{\sqrt{1+2\cosh4\varphi}},\\
		 &\omega_{\text{IV}}=\omega_{\text{IV}}(\varphi):=\omega_{\text{I}} \left[\frac{1+2\cosh4\varphi}{3}\right]^{1/4}.
	\end{align}
	\indent When $ c_-<-2 $, no real solution exists. \\
	\indent Note that the solutions  $ x_{\text{II}},x_{\text{III}}, $ and $ x_{\text{IV}} $ reduce to the same expression if the modulus $ m=k^2 \in [0,1] $ is allowed to take more general values. For example,  $ x_{\text{III}} $ reduces to $ x_{\text{II}} $ using $ \operatorname{ns}(z+\mathrm{i}K'|m)=\operatorname{sn}(\sqrt{m}z|m^{-1}) $. The complete classification of real-valued elliptic integrals expressed only using real parameters and the modulus $ \in [0,1] $ is given in Ref.~\cite{ByrdFriedman}. (Setting $ z=-t $, use 253.00 for  $ x_{\text{II}} $ and $ x_{\text{III}} $, and 259.00 for  $ x_{\text{IV}} $.)

\section{Global solution}\label{sec:connectingsols}
\begin{figure}[tb]
		\begin{center}
		(a) \\
		\includegraphics[width=0.95\linewidth]{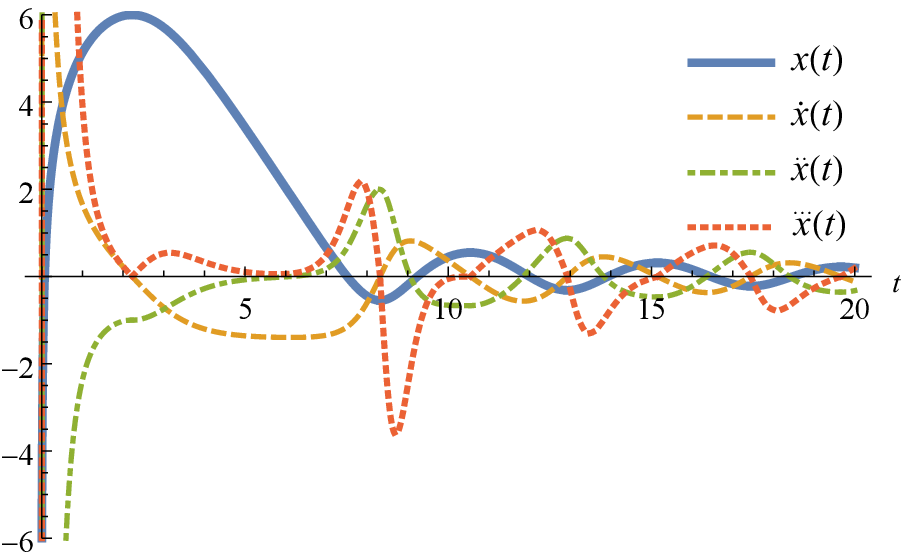} \\
		\begin{tabular}{cc}
		(b) & (c) \\
		\includegraphics[width=0.45\linewidth]{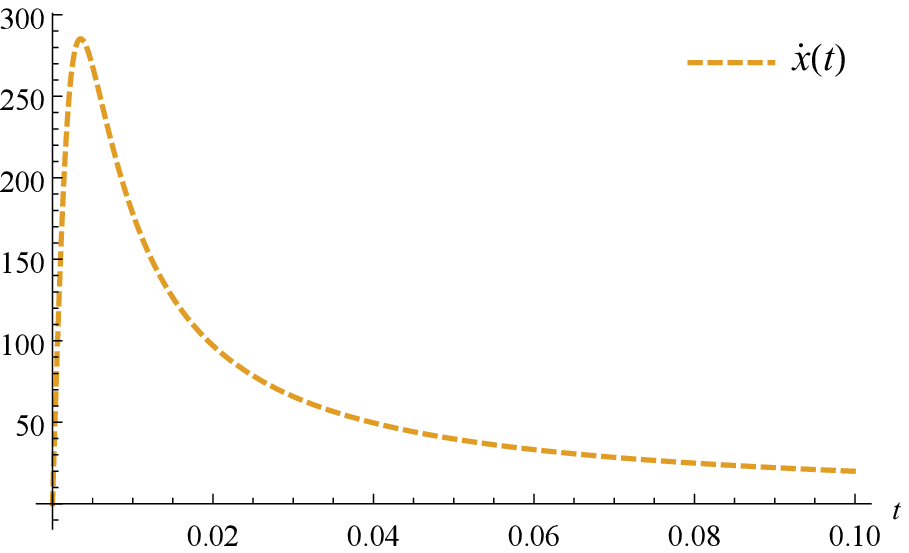} & \includegraphics[width=0.45\linewidth]{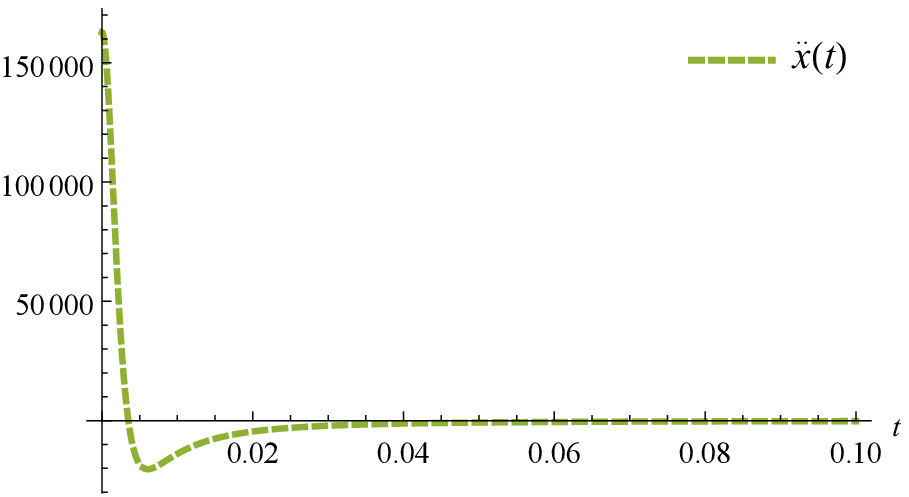} 
		\end{tabular}
		\caption{\label{fig:plotingsol}(a) The plot of the solution $ x(t) $ given by Eq.~(\ref{eq:globalsol}) and its derivatives $ \dot{x}(t),\ \ddot{x}(t) $, and $ \dddot{x}(t) $ for the Toda potential (\ref{eq:toda}). The parameters are  $ M=1,\ \kappa=1, $ and $  A=1 $. The initial condition is set to $ x(0)=-x_0=-6 $  and $ \dot{x}(0)=0 $. The first steep change in $ 0\le t \le T_{\text{I}}\simeq 2.2 $  corresponds to the fast falling in the exponential potential and climbing the linear slope. The next linear region  $ T_{\text{I}} \le t \le t_1\simeq 8.3 $ describes the uniform motion with the terminal velocity $ v_{\text{terminal}}=\sqrt{2A/(M\kappa)} $. The rest of the motion shows the damped oscillation. The cusp of the jerk $ \dddot{x}(t) $ is most visible at $ t=T_{\text{I}}\simeq 2.2 $  and $ t_1+T_{\text{I}}\simeq 10.5 $.  (b) and (c) show the the extreme behavior of $ \dot{x}(t) $ and $ \ddot{x}(t) $ at the beginning of the motion.  }
		\end{center}
	\end{figure}

	We now provide the global solution for the potential (\ref{eq:toda}) applicable for all time $ t\ge0 $, which can be constructed by smoothly matching the solutions $ x_{\text{I}},\ x_{\text{II}},\ x_{\text{III}}, $ and $  x_{\text{IV}} $ at the points $ \dot{x}(t)=0 $. For simplicity, here we consider the specific initial condition $ x(0)=-x_0<0 $ and $ \dot{x}(0)=0 $. In this case the solution is given by $ x_{\text{I}}\to x_{\text{II}} \to x_{\text{I}}\to x_{\text{II}} \to \dotsb $. When $ x(0)>0 $ and $ \dot{x}(0)=0 $, $ x_{\text{II}} $ comes first and the rest of the motion is the same. Note that $ x_{\text{III}} $ and $ x_{\text{IV}} $ are necessary only for the initial condition with very large $ \dot{x}(0)<0 $ (or very large $ x(0)<0 $ with not-so-large $ \dot{x}(0)<0 $), where the high-speed particle climbs the exponentially increasing potential slope.  They are used only once; after starting from $ x_{\text{III}} \,(\text{or }x_{\text{IV}}) $, the remaining motion is described by $ x_{\text{III}} \,(\text{or }x_{\text{IV}}) \to x_{\text{I}} \to x_{\text{II}} \to x_{\text{I}} \to x_{\text{II}} \to \dotsb $. \\
	\indent Here we show the result. The solution satisfying the initial condition $ x(0)=-x_0<0,\ \dot{x}(0)=0 $ is given by
	 \begin{align}
		x(t)=\begin{cases}
			x_{\text{I}}\left( t-t_n,x_n \right) & t_n \le t \le t_n+T_{\text{I}}, \\
			x_{\text{II}}\left( t-t_n-T_{\text{I}},\varphi_n \right) & t_n+T_{\text{I}}\le t \le t_{n+1},
		\end{cases} \label{eq:globalsol}
	\end{align}
	where $ T_{\text{I}} $,  $ t_n $, $ x_n $, and $ \varphi_n $ are defined as follows. Let us write
	\begin{align}
		T_{\text{I}}=\frac{\pi}{\omega_{\text{I}}},\quad T_{\text{II}}(\varphi)=\frac{2K(m_{\text{II}}(\varphi))}{\omega_{\text{II}}(\varphi)},
	\end{align}
	where $ \omega_{\text{I}},\ \omega_{\text{II}}(\varphi) $, and $ m_{\text{II}}(\varphi) $ are introduced in the previous section, and  $ K(m) $ is the complete elliptic integral of the first kind. Let us define $ \varphi_0 \in [0,\frac{\pi}{6}) $ by the relation $ x_0=-\frac{1}{\kappa}\ln[2\sin(\frac{\pi}{6}-\varphi_0)] $, and let us define $ \varphi_1,\ \varphi_2,\dots $ by the recurrence relation
	\begin{align}
		\sin\left( \tfrac{\pi}{6}-\varphi_{n+1} \right)=\frac{1}{4\sin\left( \tfrac{\pi}{6}+\varphi_n \right)}, \label{eq:recforphi}
	\end{align}
	and write $ x_n:=-\frac{1}{\kappa}\ln[2\sin(\frac{\pi}{6}-\varphi_n)] $. Equation (\ref{eq:recforphi}) arises from the matching condition between $ x_{\text{I}} $ and $ x_{\text{II}} $.  Then,  $ t_n $'s are given by 
	\begin{align}
		t_0=0,\quad t_n:=nT_{\text{I}}+\sum_{j=0}^{n-1}T_{\text{II}}(\varphi_j),\quad (n=1,2,3,\dots).  \label{eq:deftn}
	\end{align}
	\indent Here we briefly sketch the derivation of Eq. (\ref{eq:recforphi}) by the matching condition. Setting $ t=t_{n+1} $ in Eq.~(\ref{eq:globalsol}), we obtain $ x_{\text{I}}(0,x_{n+1})=x_{\text{II}}(T_{\text{II}}(\varphi_n),\varphi_n) \ \leftrightarrow \ x_{n+1}=-\frac{1}{\kappa}[2\sin(\frac{\pi}{6}+\varphi_n)] $. On the other hand, if we set $ t=t_{n+1}+T_{\text{I}} $, we have $ x_{\text{I}}(T_{\text{I}},x_{n+1})=x_{\text{II}}(0,\varphi_{n+1}) \ \leftrightarrow \  -x_{n+1}=-\frac{1}{\kappa}[2\sin(\frac{\pi}{6}-\varphi_{n+1})] $. Eliminating  $ x_{n+1} $ from these two equations, we obtain (\ref{eq:recforphi}). \\
	\indent The plot of the solution is given in Fig.~\ref{fig:plotingsol}.  
	Note that  $ x(t), \dot{x}(t), \ddot{x}(t) $ and $ \dddot{x}(t) $ are continuous, while the higher-order derivatives $ \frac{\mathrm{d}^n}{dt^n}x(t)\, (n\ge 4)  $ are generally discontinuous at the points $ \dot{x}=0 $, originating from the non-differentiable $ \operatorname{sgn}\dot{x} $ factor in Eq.~(\ref{eq:eom}). We can indeed observe the cusp of the jerk $ \dddot{x}(t) $ in Fig~\ref{fig:plotingsol}.

\section{Asymptotics and Envelopes}\label{sec:asym}
	\begin{figure}[tb]
		\begin{center}
		\includegraphics[width=0.95\linewidth]{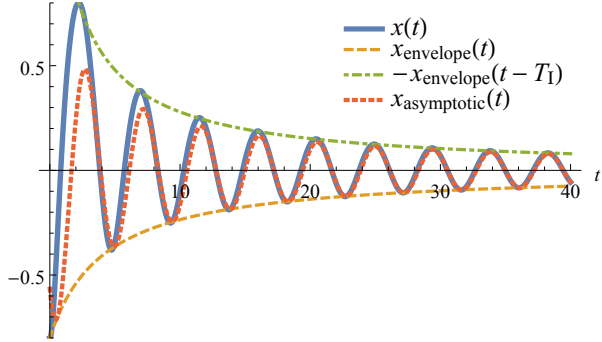}
		\caption{\label{fig:enveandasymp}The envelope (\ref{eq:envelope}) and the asymptotic curve (\ref{eq:asym}). We set the parameters  $ M=1,\ \kappa=1, $ and $  A=1 $, and the initial condition is $ x(0)=-x_0=-0.8 $ and $ \dot{x}(0)=0 $. }
		\end{center}
	\end{figure}
	Here, we determine the asymptotic behavior of the solution (\ref{eq:globalsol}). The solution of the matching condition, which is given by the recurrence relation (\ref{eq:recforphi}), is well approximated by
	\begin{align}
		\sin\left( \frac{\pi}{6}-\varphi_n \right) \simeq \frac{1}{2}\left(1-\frac{3[1-2\sin\left( \frac{\pi}{6}-\varphi_0 \right)]}{3+4n[1-2\sin\left( \frac{\pi}{6}-\varphi_0 \right)]}  \right). \label{eq:recapprosol}
	\end{align}
	It follows from that $ \delta_n:=\frac{1}{2}-\sin\left( \frac{\pi}{6}-\varphi_n \right) $ satisfies $ \delta_{n+1}-\delta_n \simeq \frac{8}{3}\delta_n^2 $. Using Eq.~(\ref{eq:recapprosol}), the approximate expression for  $ x_n=-\frac{1}{\kappa}\ln[2\sin(\frac{\pi}{6}-\varphi_n)] $ is
	\begin{align}
		x_n \simeq \frac{3x_0}{3+4n\kappa x_0}, \label{eq:xnasymp}
	\end{align}
	if the initial $ x_0 $ is not so large ($ \kappa x_0 \lesssim 1$). Within the same approximation,  $ T_{\text{II}}(\varphi_j) $ is estimated to be
	\begin{align}
		T_{\text{II}}(\varphi_n)\simeq T_{\text{I}}\left[ 1+\frac{5}{12}\left( \frac{3\kappa x_0}{3+4n\kappa x_0} \right)^2 \right].
	\end{align}
	Using this, the summation of $ t_n $ in Eq.~(\ref{eq:deftn}) can be carried out using the polygamma function. For large $ n $, it reduces to
	\begin{align}
		t_n\simeq 2nT_{\text{I}}\left[ 1+\frac{5\kappa^2 x_0^2}{8(3+4n\kappa x_0)} \right].  \label{eq:tnasymp}
	\end{align}
	Solving it with respect to $ n $, we get  $ n\simeq \frac{t_n}{2T_{\text{I}}}\left( 1-\frac{5\kappa x_0}{16}\frac{T_{\text{I}}}{t_n} \right) $, and using this, we can eliminate $ n $ from Eq.~(\ref{eq:xnasymp}). Thus we arrive at the expression for the lower envelope
	\begin{align}
		x_{\text{envelope}}(t)= -\frac{3T_{\text{I}}x_0}{3T_{\text{I}}+2\kappa x_0 t}. \label{eq:envelope}
	\end{align}
	The upper envelope is given by $ -x_{\text{envelope}}(t-T_{\text{I}}) $.\\
	\indent Next, let us derive the asymptotic curve. Within the cosine-curve approximation, the $ n $-th oscillation is given by $ \propto \cos\frac{2\pi(t-t_\infty)}{t_n-t_{n-1}} $. Up to the present accuracy, we simply obtain $ t_n-t_{n-1}\simeq 2T_{\text{I}} $. The phase shift $ t_\infty $ is determined by imposing that the argument of the cosine becomes $ 2n\pi $ at $ t=t_n $, and we have $ t_\infty=\frac{5T_{\text{I}}\kappa x_0}{16} $. Summarizing, the asymptotic curve is
	\begin{align}
		x_{\text{asymptotic}}(t)=x_{\text{envelope}}(t) \cos\left[ \pi\left( \frac{t}{T_{\text{I}}}-\frac{5\kappa x_0}{16} \right) \right]. \label{eq:asym}
	\end{align}
	The verification of these envelopes and the asymptotic curve is shown in Fig.~\ref{fig:enveandasymp}. \\
	\indent The damping by $ t^{-1} $ shown in Eq.~(\ref{eq:envelope}) is very slow compared to the system with linear drag force, where the decay occurs in an exponential way. This comes from the fact that the quadratic drag force is very weak when the particle moves slowly. 

\section{Discussion, Summary, and Future Outlook}\label{sec:summ}
	The main findings of the present paper can be summarized by the following (a) and (b):
\begin{enumerate}[(a)]
\item Determination of the family of solvable potentials (\ref{eq:familypot}) for the equation of motion with the quadratic drag term (\ref{eq:eom}).
\item Generalization of the matching method to construct the piecewise analytic solution in Ref.~\cite{1402-4896-90-8-085208} to the damped oscillators.
\end{enumerate}
The result (a) can be accomplished using the general implicit solution (\ref{eq:formalsol}) and (\ref{eq:formalsol2}) which are valid for arbitrary potentials. Moreover, the family includes the Toda potential (\ref{eq:toda}) as a limiting case. The global solution for this potential is constructed by matching the analytic solutions valid for each $ \dot{x}\gtrless0 $, and the resultant solution has a cusp in the third-order derivative. Such a matched solution with piecewise analytic property is the same as Ref.~\cite{1402-4896-90-8-085208}. The main new difference of our result is stated in (b). Namely, the value of the integration constant $ C $ in Eq.~(\ref{eq:formalsol}) changes at every junction point $ \dot{x}=0 $, where the generalized matching procedure well describes the damped oscillation in the dissipative system. This new situation increases the importance of the use of the solvable potential. Furthermore, we have also provided the asymptotic curves of the solutions by solving the matching condition, and have shown that the damping occurs by the rate proportional to~$ t^{-1} $ [Eqs. (\ref{eq:envelope}) and (\ref{eq:asym})]. \\
\indent Finally, we provide a few prospects. (i) In this paper, the exhaustive determination of solvable potentials has been made within the elementary and elliptic functions. If we want to extend the argument using more general class of functions such as hypergeometric or hyperelliptic ones, Eq.~(\ref{eq:formalsol2}) will still remain to be a starting point due to its generality. (ii) The investigation of the solutions for other solvable potentials in Eq.~(\ref{eq:familypot}) is left as a future task. Since they have local extrema and solutions for both signs of velocity becomes elliptic,  we expect more complicated classification based on their stability and boundedness. Verifying the damping behavior of the solutions [Eqs. (\ref{eq:envelope}) and (\ref{eq:asym})] for these more general potentials will be also important. For example, if we set $ d=0 $ in Eq.~(\ref{eq:familypot}), we get a potential with symmetric shape $ U(x)=\frac{A}{2\kappa}(\cosh 2\kappa x-1) $, which will be more suitable from the viewpoint of experimental realization and investigation of damping dynamics. One of the most well-known system described by Eq.~(\ref{eq:eom}) is the classical ball with air's drag. So, if we construct a bowl with the cosh-like shape or the Toda-type shape [Eq.~(\ref{eq:toda})], and consider the motion of the ball on this bowl in the atmosphere, the mathematical result in this paper will be verified experimentally.  (iii) Whether the appearance of Toda's potential has some implication to the soliton theory might also be worth considering. In general, soliton equations possess the inhomogeneous and higher-order generalizations. The former arises from the non-isospectral problem and used to describe relaxation phenomena in a nonuniform medium \cite{PhysRevLett.37.693,doi:10.1143/JPSJ.41.2141,GUPTA1979420}, and the latter is made from the family of the Lax pairs in the hierarchy \cite{FaddeevTakhtajan,CORREA20092522,TAKAHASHI2012632}, which were recently used as an effective model of density-modulated quantum condensates \cite{1402-4896-90-4-045205,PhysRevE.93.062224}. If we could invent a new dissipative effect for these equations with preserved integrability by extending the idea of the present paper, it would be useful to study damping phenomena of nonlinear wave packets.

\ack 
This work is supported by the Ministry of Education, Culture, Sports, Science (MEXT)-Supported Program for the Strategic Research Foundation at Private Universities ``Topological Science'' (Grant No. S1511006). \\

\bibliographystyle{iopart-num}
\providecommand{\newblock}{}

\end{document}